# Ten Simple Rules for making a vocabulary FAIR

**Short title:** *Rules for making a vocabulary FAIR*

Simon J D Cox (1), Alejandra N Gonzalez-Beltran (2), Barbara Magagna (3), Maria-Cristina Marinescu (4)

(1) CSIRO Land and Water, Research Way, Clayton, VIC 3168, Australia; ORCID: [0000-0002-3884-3420](#)

(2) Scientific Computing Department, Science and Technology Facilities Council, Rutherford Appleton Laboratory, Harwell Campus, Didcot, OX11 0QX, United Kingdom; ORCID: [0000-0003-3499-8262](#)

(3) Environment Agency Austria, Spittelauer Laende 5, 1090 Wien, Austria; ORCID: [0000-0003-2195-3997](#)

(4) Barcelona Supercomputing Center (BSC-CNS), Carrer de Jordi Girona 29, Barcelona 08034, Spain; ORCID: [0000-0002-6978-2974](#)

**Author contributions:**

SJDC - Conceptualization, Writing – Original Draft Preparation, Writing – Review & Editing, Resources

ANGB - Conceptualization, Writing – Original Draft Preparation, Writing – Review & Editing, Resources

BM - Conceptualization, Writing – Review & Editing

MCM - Conceptualization, Writing – Review & Editing

# Introduction

Environmental sustainability, global pandemics and other natural disasters are some of the challenges we are facing in the 21st century. Addressing these challenges involves analysing vast amounts of data from different sources, which is more effective when these sources are aggregated to find evidence-based solutions. Understanding the data, identifying the terminology that can be used to annotate them, and how they relate is a prerequisite to enable data integration.

Shared terminology is key to accurate communication and an enabler for data integration. Many organizations and disciplines have a tradition of curating lists of terms to serve various functions, particularly for annotating data or populating databases. These are often referred to as *glossaries*, and if there is a process to manage them, *'controlled-vocabularies'* ([https://www.sciencedirect.com/topics/computer-science/controlled-vocabulary](https://www.sciencedirect.com/topics/computer-science/controlled-vocabulary)). Such vocabularies were typically managed as lists or tables within text-based resources (books and manuals), or sometimes as authority-tables in databases or in spreadsheets, for use within very specific communities and applications. However, the data-integration challenge requires that common vocabularies are available for use across broader communities, supporting harmonization of datasets both within and across applications. This requires the vocabularies to be interoperable, i.e. able to exchange and use information across systems and disciplines.

The emergence of the semantic technology stack (https://www.w3.org/standards/semanticweb/), provides standardised knowledge representation languages and enables linked data (https://www.w3.org/standards/semanticweb/data). This has facilitated the development of new accessible and interoperable vocabularies, ranging from simple taxonomies and thesauri to highly expressive ontologies. In particular, the use of standard knowledge representation languages make a vocabulary not only useful for humans, but also for machines. Nevertheless, the legacy vocabularies represent the accumulated consensus of important disciplines and communities. Hence, making such vocabularies FAIR -- or Findable, Accessible, Interoperable and Reusable [1] -- is a high-value activity that can preserve the embedded domain intuition and knowledge. While controlled-vocabularies were often defined and used within small communities or organizations, FAIR vocabularies can be used in the context of much larger interconnected data and communities, and be actionable by machines. Moreover, the need to provide metadata is ubiquitous and these metadata should use FAIR vocabularies to be FAIR themselves. Hence, processes and practices are required for transitioning and adapting vocabularies, from traditional forms rooted in print technologies, to more broadly accessible modes which are available openly on-demand, as web resources. These have been demonstrated in many projects and services (e.g. [2]). Our goal here is to distill guidelines for taking an existing list of terms and converting it to a FAIR vocabulary, and present them as ten simple rules.

Vocabularies come in many forms, from simple word-lists, glossaries, hierarchical vocabularies, thesauri, taxonomies, through to axiomatized ontologies. Other sets of codes and terms used in data annotation, that may not be initially recognised as 'vocabularies', such as units of measure, lists of materials, taxa, substances, and reference systems like geologic and dynastic time-scales (which are composed of named intervals), can also be represented and distributed using many of the same tools.

In this paper we focus on one specific scenario, where:

1. there is a community requirement to be able to annotate or classify data or metadata, using agreed terms

2. a suitable vocabulary (list of terms or codes and definitions) is available, hereafter called the *legacy vocabulary*; it was created by an organisation, person or group of people that we refer as the 'content custodian', who may also be maintaining and revising it moving forward

3. the legacy vocabulary is in the form of a print document, a digital document, or in a semi-structured form such as a spreadsheet or comma-separated value file (CSV), and is not arranged and published in a way that allows remote reference to the individual terms, on an open and web-standard basis

4. no other vocabulary that is suitable for the application and acceptable to the community is published in a FAIR way either.

The 10 Simple Rules below describe how to convert that legacy vocabulary into a form enabling metadata and data annotation, using applicable standards, and also compatible with, and thus potentially able to be integrated with, related FAIR vocabularies.. Some of the rules refer explicitly to the main FAIR principles, while others are basic vocabulary prerequisites. This scenario is narrow, but common.

We provide extensive supplementary material online at https://fairvocabularies.github.io/examples/ in the form of detailed examples taken from real vocabularies that illustrate the rules. It is strongly recommended to consult these examples in order to more fully understand details of our Ten Simple Rules.

This paper is complementary to Ten Simple Rules about vocabulary development [3] and vocabulary selection [4].

# Rules

### Rule 1. Verify that the legacy-vocabulary license allows repurposing

Verify that the copyright-holder grants permission for the list of terms to be re-published as linked-data (noting that the copyright-holder is often different to the maintainer or content custodian - see Rule 2). In this context, 'linked-data' means (i) on the web, with an individual persistent web identifier per term (i.e. a HTTP (Hypertext Transfer Protocol) IRI (Internationalized Resource Identifier)) (ii) when a term IRI is dereferenced, a standards-comformant machine-readable representation of the term is returned.

If the source carries a Creative Commons license, then all the No Derivatives (ND) options (CC BY-**ND**, CC BY-NC-**ND)** are *not* ok, since you are developing a 'derivative product'.

The rest of the CC licenses (CC0, and CC BY, CC BY-SA, CC BY-NC, CC BY-NC-SA) are suitable, provided you are also willing to meet any BY (attribution), SA (share-alike) and NC (non-commercial) constraints.

If the original content uses another type of license, you must analyse it in the same way to understand if you are able to produce a derivative product, and what are the conditions for derivation. It may be necessary to contact the copyright-holder directly in order to explain what is planned and get permission.

### Rule 2. Determine the governance arrangements and custodian responsible for the legacy vocabulary

Identify the *content custodian*, which is the agent (i.e. organization or person/people) that was responsible for creating or selecting the list of terms in the legacy vocabulary. They will have expertise in the subject-matter. They may be an individual, a formal or informal committee or working group, or an official organization, such as a government agency, or learned society, and will usually be managing the vocabulary on behalf of a specified community, discipline, organization, and/or jurisdiction.

When you have identified the content custodian, it is recommended that you advise them of your plan to repurpose the legacy vocabulary as a FAIR vocabulary, to get their acknowledgement of your initiative. Enrol them in the repurposing process if possible. Find out their planned revision schedule for the legacy vocabulary, so that you can allow for this in your FAIR vocabulary maintenance plan (Rule 10).

## Rule 3.   Check minimal term definition completeness

Ensure there is at least (i) a label and (ii) a description or textual definition for each item in the list of terms. These are the minimum elements required for a useful vocabulary, and the minimum required information for creating the FAIR vocabulary (Rule 6). If definitions are missing, identify a provisional source for definitions (e.g. Wikipedia, DBpedia, WIkidata) and identify or recruit an expert group to review or provide the missing definitions and its sources - ideally more than one person to allow a quality control cycle.

The legacy vocabulary may also contain synonyms, intra-vocabulary relationships such as broader/narrower hierarchy, specified subsets, and cross-vocabulary mappings. Guidelines to encode all of these elements in a FAIR vocabulary are given in Rule 6.

## Rule 4.   Select a domain and service for the web identifiers

Choose a domain name for persistent identifier IRIs for the vocabulary items. The IRIs must be available over the lifetime of any datasets that will make use of them, so it should be planned to manage a HTTP server for this domain over a 10+ year time period. Since this is longer than many organization names and most organizational structures, domain names based on organizations are generally *not* suitable, except if they are of organizations specifically created for the purpose of managing vocabularies. As an alternative to managing a new HTTP server, consider existing open solutions for persistent identifiers such as [https://w3id.org](https://w3id.org) or [http://purl.org](http://purl.org) .

In Rule 9 we explain how the IRIs in this domain should be made resolvable, thus making the vocabulary, and its terms, accessible.

## Rule 5.   Design an identifier scheme and pattern

Choose and document the pattern for individual IRIs that identify terms in the vocabulary [5].

It may be planned to host multiple vocabularies in the same domain. In this case, a common pattern is to (i) have a distinct 'path' for each vocabulary, and then (ii) append a distinct local identifier, or 'local-id', as the last field to provide a unique IRI for each item in the vocabulary. In the case where the domain is planned to host a single vocabulary, the pattern may be simplified, with the rule for a local-id per term being the main concern.

The local-id may be an opaque code (e.g.numeric), or it may be based on the term or primary label for each item, or some other token, but must be unique in context. For vocabularies with up to a few hundred terms where the meanings do not change overtime, using a label as the basis for a unique local-id may be manageable, and can be a useful mnemonic for developers and maintainers. However, in this case it is important to consider the stability of the current label, and have a strategy for managing the item IRI if a different label becomes preferred for the same concept. For large vocabularies, or when labels may change over time, label-based unique local-ids are difficult to sustain, and numeric or opaque identifiers are more common [5].

It is recommended not to embed version information or other metadata in the identifier, since this creates challenges if the same concept persists over multiple versions or releases. It is recommended to avoid long IRI paths. In particular, paths which merely reproduce structures (e.g. hierarchies) that are recorded explicitly within the vocabulary are unnecessary. It is recommended to use slash ('/') IRIs, rather than hash ('#') IRIs. While the latter are often used in web applications, they are generally unsuitable for large vocabularies. This is because the fragment after the # is not sent to the HTTP server, so when a # IRI is requested the entire vocabulary will be returned instead of just a single term, which is often undesirable particularly for large vocabularies [6].

## Rule 6. Create a semantic-standards based vocabulary - Interoperability

Convert the vocabulary to semantic standards, using either the Simple Knowledge Organisation System (SKOS) [7,8] or the Web Ontology Language (OWL) [9,10], together with elements from other standard Resource Description Framework (RDF) vocabularies [11–13].

The table below details various technical steps and patterns for use of either SKOS or OWL to represent a vocabulary. There are a number of considerations in making a choice of one or the other of these pathways [14]:

- SKOS was designed for sets of definitions optionally arranged in a hierarchy, so nicely fits the primary scenario under consideration here: i.e. conversion of a legacy vocabulary to an RDF-based form using a semi-formal representation. SKOS includes a number of features designed to make the conversion straightforward, including synonyms, codes, subsets, and broader/narrower relationships. However, there are limitations in its logical completeness that are considered weaknesses in some applications;

- OWL supports logical axiomatization (based on description logics) for representing formal ontologies, and was designed for a much wider range of applications than the primary scenario. Nevertheless, while OWL axiomatization provides much stronger support for reasoning, it misses explicit constructs for some utility features that are built into SKOS (i.e. `skos:notation, skos:Collection, skos:altLabel, skos:related, skos:scopeNote`).

We refer the reader to other sources for more details on which one to choose, and how they could be used together [15]. Generally, you should not be overly concerned about the choice, since the type (as well as any other information) might be changed later while retaining the same IRIs, or else multiple alternative representations provided, as long as they describe the same underlying concept. **The most important feature of any implementation is that a unique IRI (Rule 5) is used to denote each distinct vocabulary term, and these IRIs can thus be used to annotate or classify data or metadata.** The detail of whether dereferencing a IRI results in a SKOS or OWL, or even some other RDF representation (or any combination of the above), is strictly a second-order concern for the use-case under consideration here - i.e. data annotation. The following table illustrates the different steps to follow to create a FAIR vocabulary relying on SKOS or OWL (for the namespace prefixes see the box) .

```
Namespace prefixes mentioned in the rule

dcterms:   http://purl.org/dc/terms/

owl:       http://www.w3.org/2002/07/owl#

rdf:       http://www.w3.org/1999/02/22-rdf-syntax-ns#

rdfs:      http://www.w3.org/2000/01/rdf-schema#

skos:      http://www.w3.org/2004/02/skos/core#
```

| **Step** | **SKOS** | **OWL** |
|---|---|---|
| **Identify terms** | Encode each vocabulary term as a `skos:Concept`, assigning an identifier as discussed in Rule 5 | Encode each vocabulary term as an `owl:Class`, or as an instance of an `owl:Class` if it is the most specific concept in the vocabulary [16]. The term identifier should be assigned as discussed in Rule 5 |
| **Encode term labels and synonyms** | Encode term names and synonyms as `skos:prefLabel` or `skos:altLabel`, respectively. | Encode labels for terms as `rdfs:label`. Synonyms can be encoded as annotations or as equivalent classes (owl:equivalentClass). |
| **Add textual definitions** | Textual definitions are encoded as `skos:definition` | Textual definitions are encoded as `rdfs:comment` |
| **Add codes and symbols** | Codes and symbols are encoded as `skos:notation` | Codes and symbols can be added as additional labels (`rdfs:label`) or a specific annotation property may be defined |
| **Add notes or comments for clarifications** | Comments can be encoded using `skos:note`. Clarifications on usage can be recorded using `skos:scopeNote` | Comments can be encoded using `rdfs:comment` |
| **Add per-item metadata, if available** | Individual vocabulary items may be annotated using standard elements such as `dcterms:creator`, `dcterms:created`, `dcterms:modified`, `dcterms:source`, `rdfs:seeAlso`. | The same metadata terms can be used in OWL, as well as `owl:versionInfo`, `rdfs:comment`, `rdfs:isDefinedBy`. Alternatively, you can adopt a solution for describing terms metadata such as the OBO Metadata Ontology (http://www.obofoundry.org/ontology/omo.html). |
| **Define the hierarchy of terms** | If broader/narrower relationships between items in the vocabulary are provided in the source document, encode these as `skos:broader`/`skos:narrower`. A narrower concept or subclass may be related to more than one broader concept or parent class, so each item may appear in more than one place in a hierarchy. | If broader/narrower relationships between items in the vocabulary are provided in the source document, encode these as `rdfs:subClassOf` (if items are `owl:Classes`). A narrower concept or subclass may be related to more than one broader concept or parent class, so each item may appear in more than one place in a hierarchy. |
| **Encode relationships between terms** | If other non-specific relationships between items in the vocabulary are provided in the source document, they may be encoded using `skos:related`. | You could use the generic term `dcterms:relation` to indicate related resources, but depending on the relationship types, it is advised to re-use, or create if they do not exist, OWL object properties with the specific required semantics. For examples, you can see the Relations Ontology (http://www.obofoundry.org/ontology/ro.html). |
| **Define subsets** | If subsets or other groupings of terms are present in the source, encode each as a `skos:Collection`. Collections may be nested. Concepts may be | To construct subsets in an OWL vocabulary you must specify a class to represent such sets and a property for their membership. (There is no equivalent option built-in to OWL) |
| | members of more than one collection. | |
| **Define the whole vocabulary** | The complete vocabulary should be encoded as a `skos:ConceptScheme`, with every skos:Concept having a `skos:inScheme` relationships to the scheme, or the top items in broader/narrower chains having a `skos:topConceptOf` relationship to the concept scheme. | The complete vocabulary should be encoded as an `owl:Ontology`, with `rdfs:isDefinedBy` relationships from the member terms. |

Note that you can use language tags for the term labels in multilingual vocabularies.

Different approaches will be required for the conversion, depending on the form of the source material.

- Where the original vocabulary is only available as a printed document, scanning, or even rekeying the essential information may be the only practical route; if available as a digital text document, you may be able to copy and paste the information

- Where the legacy vocabulary is tabulated, either fully or in part, it may be possible to identify a pattern or template from the elements of your vocabulary which will allow you to (fully or partly) automate the creation of the FAIR vocabulary. Tools such as *SKOS-Play!* or *sheet2rdf* can convert spreadsheets to RDF. Links to these, and to tools to convert many other formats to RDF are available at https://www.w3.org/wiki/ConverterToRdf.

- *qSKOS* (https://qskos.poolparty.biz/) is a useful structure- and quality-checker for SKOS vocabularies, and SKOSify (https://skosify.readthedocs.io/en/latest/) automates some conversion and cleaning operations.

- *Ontorat* [17] and ROBOT [18] can be used for generating terms, annotations and axioms of an OWL vocabulary based on ontology design patterns or templates; in addition, ROBOT has other functionality to automate ontology development workflows.

Either way, it is recommended to use an RDF/OWL or SKOS IDE (Integrated Development Environment) such as *TopBraid*, *Protégé*, *VocBench*, or *PoolParty* for data entry, or just for tidying up after an automated phase.

You should resist the urge to remediate the vocabulary during encoding. You may find things that you think are errors or potential improvements while encoding the terms, but it is recommended to create the FAIR vocabulary so that it represents the legacy vocabulary as closely as possible. The initial FAIR representation can serve as a baseline for future revisions, while being clearly anchored to an archival source. Changes to the content of the legacy and FAIR ~~legacy~~ vocabularies' are the prerogative of the content custodian identified in Rule 2.

It is common to manage a vocabulary in a single OWL or RDF file, which can be persisted using one of the standard OWL/RDF serializations (see also Rule 10). You might also store the content in some RDF or graph-based database or content management system, though details of those are beyond the scope of this paper.

## Rule 7. Add rich metadata - Reusability

Add metadata for the vocabulary, by adding metadata elements to the skos:ConceptScheme or owl:Ontology that represent the vocabulary-as-a-whole.

The description of the vocabulary must include at least:

- provenance and ownership information (citation of or links to the source, pointers to the organization or community responsible for the content),

- lifecycle information (creation and update dates, vocabulary status, pointers to the people responsible for the conversion and encoding, version information)

- preferably, an open license for users (e.g. CC0 or CC-BY)

Note that there is usually no advantage in CC-BY compared with CC0. Dereferencing the vocabulary or term IRIs will get resources that should contain attribution information, as long as they are annotated and hosted properly.

Different communities rely on metadata elements as defined by different vocabularies such as Data Catalog Vocabulary (DCAT [19]), Linked Open Vocabularies (LOV [20]), Ontology Metadata Vocabulary (OMV [21]), or the Metadata for Ontology Description and Publication Ontology (MOD [22]). OWL includes some built-in annotation properties that are applicable to OWL ontologies (e..g `owl:priorVersion, owl:backwardsCompatibleWith, owl:incompatibleWith`). The choice of which metadata vocabulary and details about mandatory requirements may be prescribed in policies of the vocabulary repository (Rule 8).

## Rule 8. Register the vocabulary - Findability

Load or register the encoded content in a vocabulary service or semantic repository, such as Research Vocabularies Australia (RVA) (https://vocabs.ardc.edu.au/) (for SKOS vocabularies) or BioPortal (https://bioportal.org) and its derivatives such as Agroportal (http://aims.fao.org/agroportal) and Ecoportal (http://ecoportal.lifewatchitaly.eu/ontologies) (for OWL ontologies and SKOS vocabularies). If you expect to be maintaining many vocabularies and you might establish your own service using one of the many software stacks available.

You should also deposit release snapshots of the vocabulary in a repository such as Zenodo (https://zenodo.org) or Dryad (https://datadryad.org/stash), or in an institutional data repository available to you. This step will assign a DOI to the vocabulary and will ensure that the vocabulary is indexed in search engines. See Rule 10 for recommendations of using a version control system, and consider that there are automated ways to store Github releases in Zenodo (with associated DOI). You may also consider registering the FAIR vocabulary as a 'standard' in FAIRsharing (https://fairsharing.org/).

Finally, engage with the community for whom the vocabulary is provided, through the subject-matter experts that created the vocabulary, the content custodians, as identified in Rule 2. They are likely to maintain a listing of community resources, which is often the first place that members of the community would look.

## Rule 9. Make the IRIs resolve - Accessibility

The HTTP server for the vocabulary domain (identified in Rule 4) must be configured so that any request for an IRI denoting a term/concept/class gets (or is re-written to get) the correct representation of the individual term from the primary service that hosts the vocabulary. If multiple serializations or representations are available, the HTTP server should support those being retrieved using standard HTTP content negotiation (using `Accept:` and `Accept-profile:` headers [23]). The IRI for the vocabulary-as-a-whole should get a suitable representation, such as a 'Landing Page' (if HTML is requested) or an RDF representation of the `skos:ConceptScheme` or `owl:Ontology` (if RDF is requested).

SPARQL [24,25] is the standard RDF query interface, so a SPARQL endpoint may be provided to support flexible queries and interactions. A link to the SPARQL endpoint should be provided on the HTML landing pages. The public endpoint should not allow SPARQL Update operations [26]. The hosting service may provide other vocabulary Application Programming Interfaces (e.g. RVA provides SISSvoc [27]). These should be clearly advertised to the user-community.

Finally a complete representation of the vocabulary should be available for download (i.e. as an RDF or OWL file).

## Rule 10. Implement a process for maintaining the FAIR vocabulary

The FAIR vocabulary should be created and maintained so that it reflects the content and updates agreed and issued by the content custodian, so it is important to obtain the maintenance schedule and versioning strategy for the vocabulary from the content custodian (Rule 2).

We recommend updating the FAIR vocabulary as soon as practical after the content custodian updates the legacy vocabulary. If the content custodian wishes to maintain the content in its original form (i.e. the legacy vocabulary), then try to arrange for alerts advising you of changes to be issued by the custodian, in order to trigger the process of update of the FAIR vocabulary. However, it may be possible to transition to an arrangement in which the FAIR vocabulary becomes the primary version or 'point of truth' for the content, though this can only be done with the consent of the content custodian.

It is common to persist the FAIR vocabulary in a single RDF file, using one of the standard RDF serializations (e.g. Turtle, RDF-XML, JSON-LD). It is strongly recommended to manage this in a version control system from the start, preferably publicly accessible (e.g. BitBucket, GitHub, GitLab). It is recommended to use an associated issue tracker to record individual changes by the content custodian, or proposals by members of the community. This allows maintenance to be transparent, and explanations of individual changes between versions.

If revision of the vocabulary is by new releases of the vocabulary-as-a-whole, then status and version information will be in the vocabulary metadata (see Rule 7). If maintenance is continuous, then item metadata should reflect this (see Rule 6). Standard Dublin Core, SKOS and OWL properties that may be useful in versioning include:

- `dcterms:created` - date or date-time that the vocabulary or term was initially created
- `dcterms:modified` - date or date-time that the vocabulary or term was last updated
- `dcterms:isReplacedBy` - to point to a superseding vocabulary or term
- `dcterms:replaces` - to point to a prior version of a vocabulary or term
- `owl:deprecated = 'true'` if the vocabulary or term is no longer valid
- `owl:priorVersion` - to point to a previous version of a vocabulary
- `owl:versionInfo` - general annotations relating to versioning
- `skos:changeNote` - modifications to a term relative to prior versions
- `skos:historyNote` - past state/use/meaning of a term

Do not re-assign or remove identifiers. If necessary, you can deprecate or retire an identifier. However, ensure that IRIs for retired and superseded items remain de-referenceable, as well as for previous versions of the vocabulary, so that references to them still return a result. .

Terms that carry over between releases without the definition changing must retain the same IRI: if the IRI changes, then datasets that use different versions of the same vocabulary cannot interoperate. Consult with the content custodian to clarify the 'identity-determining' characteristics of items, but note that changing relationships (e.g. position in a hierarchy) or the textual definition do not *necessarily* require changing the identifier (i.e. minting a new IRI) provided that the intention for the concept is still the same. It is recommended to clearly distinguish changes of 'what we currently know' about a concept, from changes to the intention or essential semantics.

# Summary and Conclusion

We have presented ten simple rules that support converting a legacy vocabulary - a list of terms available in a print-based glossary or table not accessible using web standards - into a FAIR vocabulary. Various pathways may be followed to publish the FAIR vocabulary, but we emphasise particularly the goal of providing a distinct IRI for each term or concept. A standard representation of the concept should be returned when the individual IRI is de-referenced, using SKOS or OWL serialised in an RDF-based representation for machine-interchange, or in a web-page for human consumption. Guidelines for vocabulary and item metadata are provided, as well as development and maintenance considerations.

By following these rules you can achieve the outcome of converting a legacy vocabulary into a standalone FAIR vocabulary, which can be used for unambiguous data annotation. In turn, this increases data interoperability and enables data integration. A set of examples illustrating the application of these rules are provided as supplementary material at https://fairvocabularies.github.io/examples/.

Further steps towards broader interoperability that may be considered, but are beyond the scope of this paper, include:

- relationships to terms and definitions in other FAIR vocabularies

- patterns for re-use of terms from and subsets of existing FAIR vocabularies
- supplementation of generic SKOS/OWL encoding with domain-based elements and axiomatization (see examples in the supplementary material)
- rules for maintenance

These will be addressed in future guidelines.

# Acknowledgments

We thank CODATA (https://codata.org) and the DDI Alliance (https://ddialliance.org/), who organised a Workshop on Cross-domain Metadata at Schloss Dagstuhl in October 2019, where this work was initiated.

Contributions by SJDC were supported through a CSIRO strategic project on engagement with CODATA. Contributions by BM were supported through eLTERplus, a project funded from the INFRAIA-01-2018-2019 programme of European Union's Horizon 2020 research and innovation programme under grant agreement No 871128.